\documentclass[prl,floats,aps,epsfig,nofootinbib,axodraw,amssymb,11pt]{revtex4}
\usepackage{amsfonts}
\usepackage{mathrsfs}
\usepackage{amsmath}
\usepackage{xcolor}
\usepackage{amssymb}
\usepackage{bm}
\usepackage{feynmp}
\usepackage{extarrows}
\usepackage{slashed}
\usepackage{graphicx}
\usepackage{subfigure}

\usepackage[linktocpage=true,bookmarks=true,bookmarksnumbered=true,breaklinks=true,pdfpagemode=Fullscreen,pdfstartview=FitBH]{hyperref}
\def\bea{\begin{eqnarray}}
\def\eea{\end{eqnarray}}
\def\nn{\nonumber}
\def\be{\begin{equation}}       \def\ee{\end{equation}}
\def\beq{\begin{equation}}      \def\eeq{\end{equation}}
\def\bea{\begin{eqnarray}}      \def\eea{\end{eqnarray}}
\def\beqa{\begin{eqnarray}}      \def\eeqa{\end{eqnarray}}
\def\ba{\begin{array} }
\def\ea{\end{array} }
\def\bc{\begin{center}}
\def\ec{\end{center}}
\def\beqs{\begin{subequations}}
\def\eeqs{\end{subequations}}
\def\bnum{\begin{enumerate} }
\def\enum{\end{enumerate}}
\def\nn{\nonumber}
\def\f{\frac}

\def\geqq{\geqslant}

\def\nn{\nonumber}
\def\f{\frac}

\def\[{\left[}
\def\]{\right]}
\def\({\left(}
\def\){\right)}

\def\ifb{\textrm{fb}^{-1}}

\def\Si{\Sigma}

\def\pslash{\not{\hbox{\kern-4pt $p$}}}
\def\qslash{\not{\hbox{\kern-4pt $q$}}}
\def\lv{\not{\hbox{\kern-4pt $L$}}}
\def\lsim{\mathrel{\raise.3ex\hbox{$<$\kern-.75em\lower1ex\hbox{$\sim$}}}}
\def\gsim{\mathrel{\raise.3ex\hbox{$>$\kern-.75em\lower1ex\hbox{$\sim$}}}}
\def\ifmath#1{\relax\ifmmode #1\else $#1$\fi}

\usepackage{graphicx}
\usepackage{bm}

\begin{document}

\baselineskip 17.5pt

\setcounter{page}{1}

\title{\hspace*{-8mm}Discovering New Gauge Bosons of Electroweak Symmetry Breaking at LHC-8}

\author{Chun Du\,$^1$,~ Hong-Jian He\,$^{1,2}$,~ Yu-Ping Kuang\,$^1$,~ Bin Zhang\,$^1$}
\affiliation{$^1$\,Center for High Energy Physics, Tsinghua University, %
                   Beijing 100084, China\\
             $^2$\,Theory Division, CERN, CH-1211 Geneva 23, Switzerland}

\author{Neil D. Christensen}
\affiliation{Pittsburgh Particle Physics, Astrophysics and Cosmology Center, %
Department of Physics and Astronomy, University of
Pittsburgh, Pittsburgh, PA 15260, USA}

\author{R. Sekhar Chivukula,~ Elizabeth H. Simmons}
\affiliation{Department of Physics and Astronomy, Michigan State University,
East Lansing, Michigan 48824, USA}


\begin{abstract}
\baselineskip 17.5pt
We study the physics potential of the 8\,TeV LHC (LHC-8) to discover, during
its 2012 run, a large class of extended gauge models or extra dimensional
models whose low energy behavior is well represented by an
$SU(2)^2 \otimes U(1)$ gauge structure.
We analyze this class of models and find that with a combined integrated
luminosity of $40-60$\,fb$^{-1}$ at the LHC-8, the first new
Kaluza-Klein mode of the $W$ gauge boson can be discovered up to a
mass of about $370 - 400$\,GeV, when produced in association with a $Z$ boson.
\\[1mm]
\hspace*{-7mm}
{PACS numbers:\,12.60.Cn,\,11.10.Kk,\,12.15.Ji,\,13.85.Qk}.~~  
Phys.\,Rev.\,D\,(2012) [\,{\tt arXiv:1206.6022}\,]
\\[-1mm]
\hspace*{-7mm}
{Preprint$^{\#}$: {\small CERN-PH-TH/2012-177,
MSUHEP-120618, PITT-PACC-1206, TUHEP-TH-12177}}
\end{abstract}

\maketitle

\baselineskip 18pt

\vspace*{5mm}
\section{1.~Introduction}

By the end of 2011 the LHC, running at a center of mass energy of 7
TeV, had accumulated an integrated luminosity of about 5\,$\ifb$
from both the ATLAS and CMS experiments \cite{LHC7-2011}. Since
April 5, 2012, the LHC has been running at an 8\,TeV collision
energy, and has collected about 12\,$\ifb$ of data in each
detector by August 20. The LHC, running in this ``LHC-8'' mode, is expected to
produce up to about $\,20-30\,\ifb$\, of data apiece in the ATLAS
and CMS detectors by the end of this year, which will amount to
$40-60\,\ifb$ in total.  This will enable the LHC to make incisive
tests of the predictions of many competing models of the origin of
electroweak symmetry breaking (EWSB), from the Standard Model (SM)
with a single Higgs boson, to models with multiple Higgs bosons, and
to so-called Higgsless models of the EWSB.
The Higgsless models \cite{Csaki:2003dt}
contain new spin-1 gauge bosons which
play a key role in EWSB by delaying unitarity violation
of longitudinal weak boson scattering up to a higher ultraviolet (UV)
scale \cite{SekharChivukula:2001hz} without invoking a fundamental
Higgs scalar. Very recently, the effective UV completion of the minimal
three-site Higgsless model \cite{3site} was presented and studied
in \cite{Abe:2012fb} which showed that the latest LHC signals of a
Higgs-like state with mass around $125-126$\,GeV \cite{July4}
can be readily explained, in addition to the signals of new spin-1
gauge bosons studied in the present paper.

In this work, we explore the physics potential of the LHC-8 to discover a
relatively light fermiophobic electroweak gauge boson $W_1$ with
mass $250-400$\,GeV, as predicted
by the minimal three-site moose model\,\cite{3site} and
its UV completion\,\cite{Abe:2012fb}.  Being fermiophobic or nearly so,
the $W_1$ state is allowed to be fairly light. More specifically,
the 5d models that incorporate ideally
\cite{SekharChivukula:2005xm} delocalized fermions
\cite{Cacciapaglia:2004rb,Foadi:2004ps}, in which the ordinary
fermions propagate appropriately in the compactified extra dimension
(or in deconstructed language, derive their weak properties from
more than one $SU(2)$ group in the extended electroweak sector
\cite{Chivukula:2005bn,Casalbuoni:2005rs}), yield phenomenologically
acceptable values for all $Z$-pole observables \cite{3site}. In this
case, the leading deviations from the SM appear in
multi-gauge-boson couplings, rather than the oblique parameters
$S$ and $T$. Ref.\,\cite{tri-3site} demonstrates that the LEP-II
constraints on the strength of the coupling of the $Z_0^{}$-$W_0^{}$-$W_0^{}$
vertex allow a $W_1$ mass as light as 250\,GeV,
where $W_0$ and $Z_0$ refer to the usual electroweak gauge bosons.

In the next section we introduce the model.  Section\,3
presents our analysis of the $\,pp\rightarrow W_1Z_0\rightarrow
W_0Z_0Z_0 \rightarrow jj \ell^+ \ell^-\ell^+ \ell^-$ process at the
LHC-8. Finally, we demonstrate that the LHC-8 should be able to sensitively
probe $W_1$ bosons in the mass-range of $250-400$\,GeV by the end of
this year.

\section{2.~The Model}

\begin{figure}[t]
\begin{center}
\includegraphics[scale=1.2]{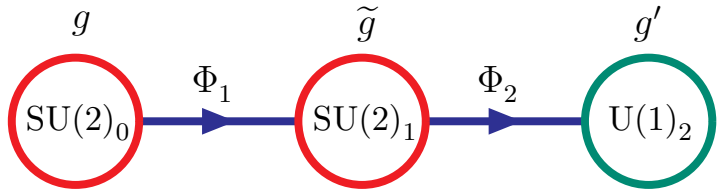}
\end{center}
\null\vspace{-1cm}
\caption{Moose diagram of the minimal linear moose model (MLMM)
with the gauge structure $SU(2)_0 \times SU(2)_1 \times U(1)_2$\,
as well as two independent link fields $\Phi_1^{}$ and $\Phi_2^{}$
for spontaneous symmetry breaking.
The relevant parameter space of phenomenological interest
is where the gauge couplings obey $\,g,g' \ll \tilde{g}$\,.}
\label{fig:models}
\end{figure}

We study the minimal deconstructed moose model at LHC-8
in a limit where its gauge sector is equivalent to the  ``three site model" \cite{3site}
or its UV completed ``minimal linear moose model" (MLMM)\,\cite{Abe:2012fb}
whose gauge boson phenomenology was previously studied \cite{PRD2008}\cite{3siteLHC-ex}
for the 14\,TeV LHC.
Both the three site model and the MLMM are based on the gauge group
$\,SU(2)_0\otimes SU(2)_1\otimes U(1)_2$,
as depicted by Fig.\,\ref{fig:models} and its
gauge sector is the same as that of the BESS models \cite{Casalbuoni:1985kq,Casalbuoni:1996qt} 
or the hidden local symmetry model
\cite{Bando:1985ej,Bando:1985rf,Bando:1988ym,Bando:1988br,Harada:2003jx}.
The extended electroweak symmetry spontaneously breaks to electromagnetism
when the distinct Higgs link-fields $\Phi_1$ connecting $SU(2)_0$ to
$SU(2)_1$ and $\Phi_2$ connecting $SU(2)_1$ to $U(1)_2$ acquire
vacuum expectation values (VEVs) $\,f_{1}^{}$ and $\,f_{2}^{}$\,.\,
The weak scale $\,v \simeq 246$\,GeV is related to those VEVs via
$\,v^{-2} = f_1^{-2} + f_2^{-2}$\, %
and, for illustration, we take $\,f_1^{}=f_2^{}=\sqrt{2}v$\,.\,
Below the symmetry breaking scale, the gauge boson spectrum includes an extra
set of weak bosons $(W_1^{},\,Z_1^{})$,\, in addition to
the standard-model-like weak bosons $(W_0^{},\,Z_0^{})$ and the photon.
Furthermore, the scalar sector of the MLMM\,\cite{Abe:2012fb}
contains two neutral physical Higgs bosons
$(h^0,\,H^0)$, as well as the six would-be Goldstones eaten by
the corresponding gauge bosons $(W_0^{},\,Z_0^{})$ and $(W_1^{},\,Z_1^{})$.

In our previous work \cite{PRD2008} on the phenomenology of
such spin-1 new gauge bosons at a 14\,TeV LHC,
we studied the potential for detecting the
$W_1^{}$ via both the weak boson fusion $\,pp\rightarrow
W_0^{}Z_0^{}jj \rightarrow W_1^{} jj \to W_0^{} Z_0^{} jj$\, and the
associated production process $\,pp\rightarrow
W_1^{}Z_0^{}\rightarrow W_0^{}Z_0^{}Z_0^{}$\,.\, Focusing on the
mass range $400-1000$\,GeV, we found that the associated production
would require less integrated luminosity than the gauge boson fusion
channel at the lower end of that mass range, as shown in Fig.\,4 of
Ref.\,\cite{PRD2008}. Extrapolating that result to lower $W_1^{}$
masses and a lower LHC collision energy, we have found in this work
that for the LHC-8, the best process for detecting  $W_1^{}$ in the
mass range $250-400$\,GeV is also the associated production,
$\,pp\rightarrow W_1^{}Z_0^{}\rightarrow W_0^{}Z_0^{}Z_0^{}
 \rightarrow jj \ell^+ \ell^-\ell^+ \ell^-$,\,
where we select the $W_0^{}$ decays into dijets and the $Z_0^{}$ decays into
electron or muon pairs.

One distinctive feature of the MLMM is that the unitarity of high-energy longitudinal
weak boson scattering is maintained jointly by the exchange of both the new
spin-1 weak bosons and the spin-0 Higgs bosons \cite{Abe:2012fb}.
This differs from either the SM (in which unitarity of longitudinal weak boson scattering
is ensured by the exchange of the Higgs boson alone)\,\cite{SMuni} or the conventional
Higgsless models (in which unitarity of longitudinal weak boson scattering
is ensured by the exchange of spin-1 new gauge bosons alone)\,\cite{SekharChivukula:2001hz}.
It has been shown \cite{tri-3site} that the scattering amplitudes
in such highly deconstructed models with only three sites
can accurately reproduce many aspects of the low-energy behavior
of 5d continuum theories.

The original Lagrangian of the three site model
was given in a nonlinear Higgsless form \cite{3site},
\begin{eqnarray}
\label{eq:TMHLM}
\mathscr{L}_{\textrm{HL}}^{} &=&
\frac{1}{4}\mathrm{Tr}
\left[f_1^{2}(D_{\mu}\Si_{1})^{\dagger}(D^{\mu}\Si_{1}^{})
      + f_2^{2}(D_{\mu}\Si_{2})^{\dagger}(D^{\mu}\Si_{2}^{})\right] ,
\end{eqnarray}
where the nonlinear sigma fields
$\,\Si_j = \exp[i\pi_j^a\tau^a/f_j^{}]$\, and $\tau^a$ denotes the Pauli matrices.
The gauge covariant derivatives take the following forms,
\beqs
\begin{eqnarray}
D^\mu\Si_1^{} &=&
\partial^\mu\Si_1^{} + igW_L^{a\mu}\frac{\tau^a}{2}\Si_1^{}
-i\tilde{g}\Si_1^{}W_H^{a\mu}\frac{\tau^a}{2}\,,
\\
D^\mu\Si_2 &=&
\partial^\mu\Si_2^{} + i\tilde{g}W_H^{a\mu}\frac{\tau^a}{2}\Si_2^{}
 -i{g'}\Si_2^{}W_R^{3\mu}\frac{\tau^3}{2}\,.
\end{eqnarray}
\eeqs
Extending this construction,
we will include the radial Higgs excitations in the sigma fields.
We introduce the two radial Higgs excitations $\,h_j^{}\,$ 
as follows,
\beqa
\Phi_j^{} ~=~ (f_j^{}+h_j^{})\Si_j^{}\,, ~~~~~~
\Si_j^{} ~=~ \exp[i\pi_j^a\tau^a/f_j^{}] \,,
\eeqa
where the Higgs fields $\,\Phi_j^{}\,$ are $2\times 2$ matrices,
and the Higgs bosons $\,h_{1,2}^{}\,$ are gauge-singlets.
Thus, we can write down the Lagrangian of the MLMM by
including the radial Higgs excitations for (\ref{eq:TMHLM}),
\begin{eqnarray}
\label{eq:TMHLM-h}
\mathscr{L} &=&
\frac{1}{4}\mathrm{Tr}
\left[(D_{\mu}\Phi_1^{})^{\dagger}(D^{\mu}\Phi_1^{})
      +(D_{\mu}\Phi_2^{})^{\dagger}(D^{\mu}\Phi_2^{})\right]
- V(\Phi_1^{},\Phi_2^{}) \,,
\end{eqnarray}
where $\,V(\Phi_1^{},\Phi_2^{})\,$
denotes the scalar potential as given in \cite{Abe:2012fb},
but is not needed for the current study.
In the unitary gauge, this Lagrangian
is identical to the renormalizable MLMM studied in \cite{Abe:2012fb}.
Since our current phenomenological study (next section) just focuses on
the detection of spin-1 new gauge bosons in the MLMM,
the radial Higgs excitations included in the Lagrangian (\ref{eq:TMHLM-h})
do not affect our collider analysis.
For the following LHC analyses, we will always take
$\,f_1^{}=f_2^{}=\sqrt{2}v$\,.\,

The unitarity of the generic longitudinal scattering amplitude
of $\,W_{0}^{L}W_{0}^{L}\to W_{0}^{L}W_{0}^{L}$,\, in the presence of any numbers
of spin-1 new gauge bosons $V_k^{}\,(=W_k^{},Z_k^{})$
and spin-0 Higgs bosons $h_k$, was recently studied in Ref.\,\cite{Abe:2012fb}.
It has been shown that requiring the exact cancellation of the asymptotic $E^2$
terms\footnote{Here $E$ denotes the center-of-mass energy of the relevant scattering process.}
in the scattering amplitude imposes the following sum rule
on the couplings and masses \cite{Abe:2012fb},
\beqa
\label{eq:SR}
&&\hspace*{-5mm}
G_{4W_0}^{} -\f{3M_{Z_0}^2}{4M_{W_0}^2}G_{W_0W_0Z_0}^2 \,=\,
\sum_k \f{3M_{Z_k}^2}{4M_{W_0}^2}G_{W_0W_0Z_k}^2 +
\sum_k\f{\,G_{W_0W_0h_k}^2\,}{4M_{W_0}^2} \,.
\eeqa
Here $\,G_{{V_i}{V_j}{V_k}}$\, is the cubic coupling
among the three vector bosons indicated,
$Z_k^{}$ is the kth Kaluza-Klein mode of the $Z$ boson, and
$G_{4{W_0}}$ is the quartic coupling of $W_0^{}$ bosons.
Eq.\,(\ref{eq:SR}) extends the corresponding Higgsless sum rule
derived in \cite{SekharChivukula:2008mj}.
For the current MLMM, the general sum rule (\ref{eq:SR})
becomes \cite{Abe:2012fb},
%
\beqa
\label{eq:SR-MLMM}
G_{4W_0}^{} - \f{3M_{Z_0}^2}{4M_{W_0}^2}G_{W_0W_0Z_0}^2 \,=\,
\f{3M_{Z_1}^2}{4M_{W_0}^2}G_{W_0W_0Z_1}^2 +
\f{\,G_{W_0W_0h}^2\!+G_{W_0W_0H}^2\,}{4M_{W_0}^2} \,,
\eeqa
%
where the symbols $(h,\,H)$ denote the two mass-eigenstate Higgs bosons, and
we have $\,G_{W_0W_0h_1}^2\!+G_{W_0W_0h_2}^2=G_{W_0W_0h}^2\!+G_{W_0W_0H}^2\,$.\,
Because there is only a single extra set of weak gauge bosons in
this theory, the sum over KK modes on the right-hand-side of
(\ref{eq:SR}) reduces to a single term.
Then, with the Lagrangian (\ref{eq:TMHLM-h}) of the MLMM,
we have explicitly verified the sum rule (\ref{eq:SR-MLMM}).
Hence, the unitarity of the longitudinal
weak boson scattering in the MLMM is ensured jointly\,\cite{Abe:2012fb} by
exchanging both the new spin-1 weak bosons $W_1/Z_1$ and the spin-0 Higgs bosons $h/H$.\,
We also note that the $hWW$ and $hZZ$ couplings are generally suppressed\,\cite{Abe:2012fb}
relative to the SM values because of the VEV ratio $\,f_2^{}/f_1^{}=O(1)\,$
and the $h-H$ mixing. As shown in \cite{Abe:2012fb}, the MLMM
can predict an enhanced diphoton rate for the light Higgs boson $h$ with mass
$125-126$\,GeV via gluon fusions,
while the Higgs signals via the associate production
$\,q\bar{q}'\to hV_0\,$ and vector boson fusion
$\,q\bar{q}'\to hq_3^{}q_4^{}\,$ (with $h\to b\bar{b},\,\tau\bar{\tau}$)
are always lower than the SM.

\section{3.~Analysis of $\,{\bf W}_{\bf 1}^{\mathbf{\pm}}$ Detection at the LHC-8}

In this section, we study the partonic-level signals
and backgrounds for detecting $W_1^\pm$ states
at the LHC-8 in the associated production channel.
The signal events proceed via the process
$\,pp\rightarrow W_1Z_0\rightarrow W_0Z_0Z_0 \rightarrow jj \ell^+ \ell^-\ell^+ \ell^-$
where the leptons can be either electrons or muons\,
We have systematically computed all the major SM backgrounds
for the $jj4\ell$ final state,
including the irreducible backgrounds $\,pp\to W_0Z_0Z_0\to jj4\ell\,$ ($jj=qq'$)
without the contribution of $W_1$,\,
as well as the reducible backgrounds
 $\,pp\to ggZ_0Z_0\to jj4\ell$\,,\,
$\,pp\to Z_0Z_0Z_0\to jj4\ell\,$, and the SM $\,pp\to jj4\ell\,$
other than the above reducible backgrounds.

We performed the parton level calculations at tree-level using two
different methods and two different gauges to check the consistency.
In one calculation, we used the helicity amplitude approach
\cite{helicity} to generate the signal and backgrounds. We also
calculated both the signal and background using CalcHEP
\cite{Pukhov:1999gg,Pukhov:2004ca}. For the signal calculation in
CalcHEP, we used FeynRules \cite{Christensen:2008py} to implement
the minimal Higgsless model \cite{Christensen:2009jx}. We found
satisfactory agreement between these two approaches and between both
unitary and 't\,Hooft-Feynman gauge. We used a scale of
$\,\sqrt{\hat s}$\, for the strong coupling in the backgrounds and
$\,\sqrt{\hat s}/2$\, for the CTEQ6L \cite{cteq6} parton
distribution functions.  We included both the first and second
generation quarks in the protons and jets, and both electrons and
muons in the final-state leptons.
\begin{figure}[t]
\vspace*{-4mm}
\includegraphics[width=12.5truecm,clip=true]{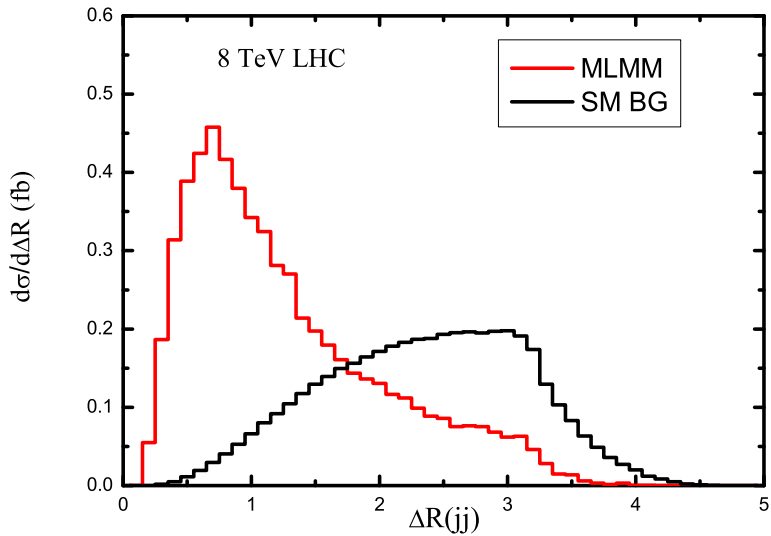}
\vspace*{-4mm}
 \caption{Event distribution $\,\Delta R(jj)$\, at LHC-8,
 for the MLMM with $\,M_{W1}^{}=300\,$GeV (red curve),
 and for the SM backgrounds (black curve) which peak around the large $\,\Delta R(jj)$.}
\label{Fig:1}
\end{figure}

In our calculations, we impose basic acceptance cuts,
\begin{eqnarray}
 p_{T \ell}^{} > 10\,{\rm GeV},  ~~~~~  |\eta_\ell^{}|<2.5 \,,
\nn\\[0.9mm]
 p_{T j}^{} > 15\,{\rm GeV},  ~~~~~  |\eta_j^{}|<4.5 \,,
\end{eqnarray}
and also a reconstruction cut for identifying $W_0$ bosons that
decay to dijets,
\begin{eqnarray}
 M_{jj}^{} = 80 \pm 15\,{\rm GeV} \,.
\end{eqnarray}
The same cuts were imposed for our previous analysis for the
14\,TeV LHC  \cite{PRD2008}, where we found that a minimum separation cut
on the two jets was not necessary.  We find that these cuts are also effective
for $W_1^\pm$ searches at the LHC-8.

 \begin{figure}[t]
 \vspace*{-5mm}
  \centering
   \includegraphics[width=12.5truecm]{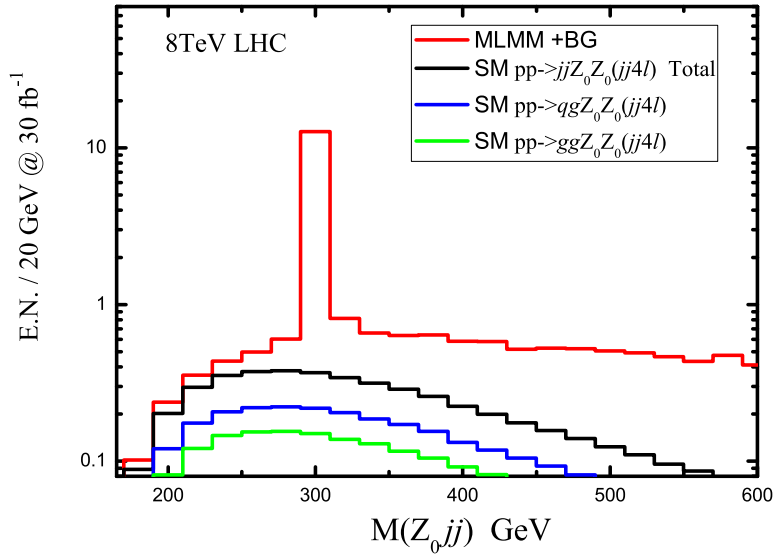}%
  \vspace*{-5mm}
  \caption{Invariant-mass distribution of \,$M(Z_0jj)$\, for
 the predicted signals of the $W_1^\pm$ bosons with mass $\,M_{W1}=300$\,GeV
 and after all relevant cuts.
 The key of this plot identifies all curves in the order from top to bottom.}
 \label{Fig:2}
\end{figure}

We further analyze the distributions of the dijet opening-angle
$\,\Delta R(jj)$\, in the decays of $\,W_0\to jj$\, for both the
signal and SM background events. This is depicted in
Fig.\,\ref{Fig:1}. We find that the signal events are peaked in the
small opening-angle region around $\,\Delta R(jj)\sim 0.6$\,,\,
while the SM backgrounds tend to populate the range of larger
opening angles, with a broad bump around $\,\Delta
R(jj)=1.5-3.3$\,.\, In order to sufficiently suppress the SM
backgrounds, we find the following opening-angle
cut\footnote{These are somewhat weaker than the cut of %
$\,\Delta R(jj)< 1.5\,$ imposed in \cite{PRD2008}.} to be very
effective \cite{j-separation},
\beqa
\label{eq:DR-cut1}
\Delta R(jj) ~<~ 1.6\,.
\eeqa
At the LHC-8, we note that the above cut reduces the signal events
by only $10-15\%$, but removes about $72-80\%$ of the SM backgrounds.

Next,  we present the invariant-mass distribution $M(Z_0^{}jj)$
in Fig.\,\ref{Fig:2}, where we compare the signal with all relevant SM backgrounds.
We have used $M_{W1}=300$\,GeV as a sample value for a relatively
light $W_1^{}$ boson. Because the two $Z^0$ bosons are
indistinguishable, each event is included twice, i.e.,
at the two $M(Z_0^{}jj)$ values corresponding to combining each
$Z_0$ boson with the dijets. The predicted signal events (plus SM
backgrounds and the signal-derived combinatorial background) are
shown for the MLMM (red curve). We have systematically computed
all the major SM backgrounds for the $jj4\ell$ final state,
including the irreducible backgrounds $\,pp\to W_0Z_0Z_0\to
jj4\ell\,$ ($jj=qq'$) without the contribution of $W_1$,\, as well
as the reducible backgrounds $\,pp\to qgZ_0Z_0\to
jj4\ell\,$\,(purple curve, 2nd from bottom),\, $\,pp\to ggZ_0Z_0\to
jj4\ell$\,(green curve, bottom),\, $\,pp\to Z_0Z_0Z_0\to jj4\ell\,$
and other SM processes of the form $pp\to jj4\ell$. The summed total
SM backgrounds are shown as the black curve (third from bottom) in
Fig.\,\ref{Fig:2}. The irreducible background and the reducible
backgrounds from $pp\to Z_0Z_0Z_0\to jj4\ell$ and other SM $pp\to
jj4\ell$ processes are so small that they are invisible in Fig.\,\ref{Fig:2}.
From Fig.\,\ref{Fig:2}, we see that at the LHC-8, the $W_1$ resonance peak
is distinct and the SM backgrounds are effectively suppressed.
We also note that the process
$\,pp \to W_0^* \to W_0 h^* \to W_0Z_0Z_0\to jj4\ell$\, is highly suppressed
after all the cuts including (\ref{eq:cut-Zjj}) below,
and is fully negligible in this analysis.
For the light Higgs boson $h^0$ with mass around $125-126$\,GeV, the
heavy gauge boson $W_1$ has a new decay channel $\,W_1\to W_0h\,$,\, and
its decay width relies on the Higgs mixing angle $\,\alpha\,$.\,
But it was found \cite{Abe:2012fb} that for our model setup of
$\,f_1^{}=f_2^{}$\,,\, the decay branching fraction of $\,W_1\to W_0h\,$
is fully negligible relative to that of $\,W_1\to W_0Z_0\,$
when the $\,h\to\gamma\gamma\,$ signals are consistent
with the current LHC data \cite{July4}.

\begin{figure}[t]
  \centering
   \includegraphics[width=11cm]{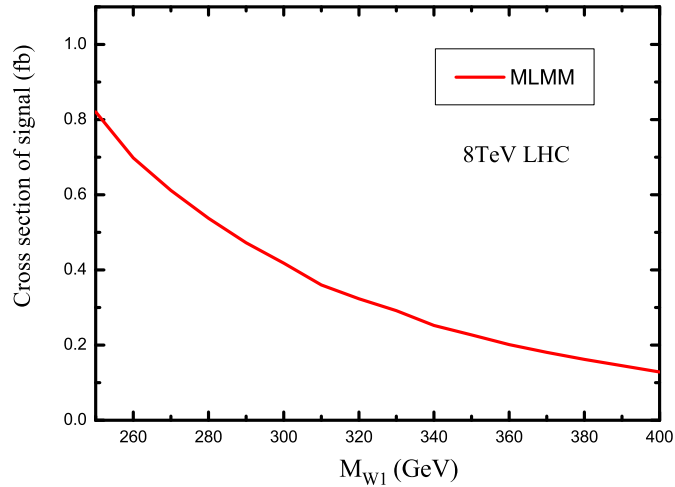}
\null\vspace{-5mm}
  \caption{Predicted signal cross section for
 $\,pp\to W_1Z_0\to W_0Z_0Z_0\to jj4\ell\,$ as a function of the
 $W_1^{}$ mass in the MLMM after all cuts at the LHC-8.}
 \label{Fig:3}
\end{figure}

\begin{table}[h]
\label{tab:1}
\begin{center}
\caption{Predicted signal cross sections of the MLMM
and the SM backgrounds for
\,$W_1^\pm$\, production via $\,pp\to W_1Z_0\to W_0Z_0Z_0\to
jj4\ell\,$ at the LHC-8, including all cuts described in the text.}
\tabcolsep0.2in
\arrayrulewidth0.5pt
\doublerulesep0.5pt
\begin{tabular}{||c|c||c|c|c|c||}
\hline\hline\hline\hline
\multicolumn{2}{||c||}{$M_{W_{1}}$ (GeV)} & 250 & 300 &350 & 400\\
\hline\hline\hline
\multicolumn{2}{||c||}{Signal Cross Section (fb)}
 & 0.8205 & 0.4180 & 0.2271 & 0.1282 \\ \hline\hline
 & $pp\rightarrow qgZ_{0}Z_{0}$ & 0.0145 & 0.0141 & 0.0114 & 0.0083  \\ \cline{2-6}
Background Cross Sections (fb)& $pp\rightarrow ggZ_{0}Z_{0}$
& 0.0101 & 0.0096 & 0.0078 & 0.0058  \\ \cline{2-6}
& Total & 0.0246 & 0.0236 & 0.0191 & 0.0141  \\
\hline\hline\hline\hline
\end{tabular}
\end{center}
\end{table}

In Fig.\,\ref{Fig:3}, we display the predicted total signal cross section
for the process $\,pp\to W_0Z_0Z_0\to jj4\ell\,$
after all cuts at the LHC-8 have been imposed; this is shown as a function of the $W_1^{}$
mass for the range $\,250-400$\,GeV.\,
Here, we define the signal region to include all events satisfying the condition,
\begin{eqnarray}
\label{eq:cut-Zjj}
 M({Z_0jj}) ~= ~M_{W_1}^{} \pm 20 \,\text{GeV} \,.
\end{eqnarray}

The cross sections of signals and backgrounds are also listed in
Table\,I for four sample values of $W_1$ masses,
$\,M_{W_1}^{}=250,\,300,\,350,\,400\,$GeV.

\section{4.~Results and Conclusions}

Finally, we present the LHC-8 discovery reach for the the relatively
light $W_1$ mass-range of $\,250-400\,$GeV. To calculate the
statistical significance, we use the Poisson probability which
governs the random generation of uncorrelated events. If the number
of events expected in the background is $\,\nu$\,,\,  then the
probability $P_P^{}(n,\nu)$ that the number of events measured will
fluctuate up to $\,n\,$ is given by
\begin{equation}
P_P^{}(n,\nu) ~=~ \frac{\,\nu^n e^{-\nu}}{n!} \,.
\end{equation}
The probability that the background will fluctuate up to the
background plus the signal or higher is then given by
\begin{equation}
P_P^{}(n\geqq \nu+s, \nu) ~=\, \sum_{n=\nu+s}^{\infty}\frac{\,\nu^n
e^{-\nu}}{n!} \,.
\end{equation}
For this to correspond to a $\,3\sigma$\, or $\,5\sigma$\,
significance, this probability must be the same as the probability
for a Gaussian to fluctuate up 3 or 5 standard deviations,
respectively, $\,P_G^{}(3\sigma)=0.00135$\, or
$\,P_G^{}(5\sigma)=2.87\times10^{-7}$\, \cite{PDG}.

In Fig.\,\ref{Fig:5}, we display the required integrated
luminosities for detecting the $W_1^\pm$ signal at the $3\sigma$ and
$5\sigma$ levels as a function of the $W_1^\pm$ mass $M_{W_1}^{}$.
Table\,II summarizes the $5\sigma$ reach in  $\,M_{W_1}$\,
for some sample values of the integrated
luminosity at the LHC-8. In this analysis, we have included
statistical errors in determining the $W_1^\pm$ discovery potential.
We anticipate that experimental analyses will include
more complete detector level simulations,
systematic errors and the details of detector geometry.

\begin{figure}[t]
\hspace*{-9mm}%
  \centering
   \includegraphics[width=12cm]{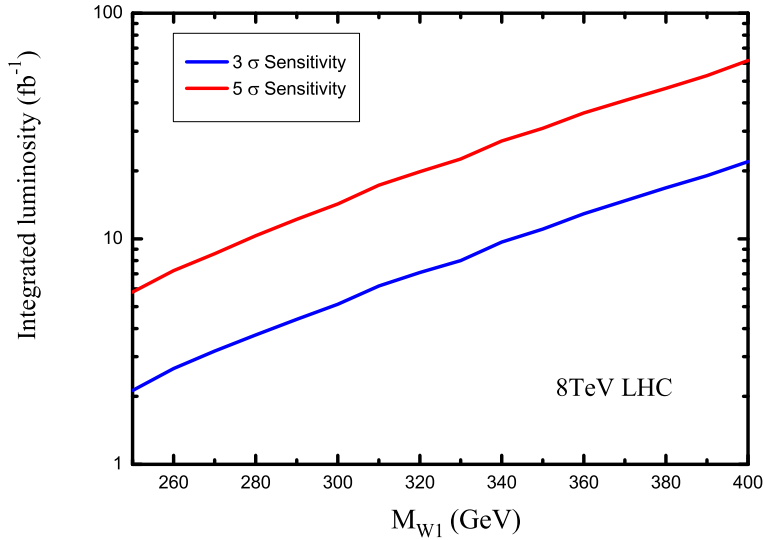}
   \vspace*{-7mm}
  \caption{Integrated luminosities required for detection of
  new $W_1^\pm$ gauge bosons at the $3\sigma$ level in the MLMM (lower blue curve),
  and at the $5\sigma$ level (upper red curve) as a function of the $W_1^{}$
  mass, at the LHC-8.} %
\label{Fig:5} %
\end{figure}

Fig.\,\ref{Fig:5} and Table\,II demonstrate that the LHC-8 should be
able to probe the light mass range for the $W_1^\pm$ gauge bosons
quite effectively in the minimal linear moose model studied here. In
fact, it has good potential for detecting $W_1^\pm$ with a mass
below 400\,GeV by the end of 2012.  This is  complementary to the
discovery reach for heavier $W_1^\pm$ bosons (400\,GeV -- 1\,TeV)
that our previous study\,\cite{PRD2008} showed to be feasible for
the LHC when running at 14\,TeV collision energy.

\begin{table}[h]
\label{tab:2}
\vspace*{-3mm} %
\begin{center}
\caption{The $5\sigma$ discovery reaches of the $W_1^\pm$ bosons at
the LHC-8, with the integrated luminosities
$\,\int\!\mathscr{L}=10,\,15,\,20,\,25,\,30,\,35,\,40,\,50,\,60$\,fb$^{-1}$,\,
respectively.} %
\vspace*{-1mm} %
\tabcolsep0.2in
\arrayrulewidth0.5pt
\doublerulesep0.5pt
\begin{tabular}{||c||c|c|c|c|c|c|c|c|c||}
\hline\hline\hline\hline
&&&&&&&&&\\[-3mm]
$\int\!\mathscr{L}$ (fb$^{-1})$ & 10 & 15 & 20 & 25 & 30 & 35 & 40 & 50 & 60
\\
&&&&&&&&&\\[-3mm]
\hline\hline
&&&&&&&&&\\[-3mm]
$M_{W_{1}}$\,(GeV)  & 277  & 302  &  320 & 335 &  346 & 357 & 367 & 385 & 397
\\[2mm]
\hline\hline\hline\hline
\end{tabular}
\end{center}
\vspace*{-1.5mm} %
\end{table}

In summary, the LHC-8 is continuing to test the origin of electroweak symmetry breaking.
The minimally extended electroweak gauge structure of $SU(2)^2\otimes U(1)$ generically
predicts the extra spin-1 gauge bosons as the unambiguous new physics beyond the SM,
which give distinct new signatures at the LHC.
We have demonstrated that after each of the ATLAS and CMS detectors collects up to
$\,20-30\,\ifb$\, of data by the end of this year, the LHC-8 should have
good potential to probe the dynamics of the extended gauge symmetry breaking
$\,SU(2)^2\otimes U(1)\to U(1)_{\text{em}}^{}$.\,
We look forward to seeing the results.

\vspace*{6mm} %
\noindent %
{\bf Acknowledgments}
\\[1mm]
This research was supported by the NSF of China (grants 11275101, 10625522,
10635030, 11135003, 11075086) and the National Basic Research
Program of China (grant 2010CB833000); by the U.S. NSF under Grants
PHY-0854889 and PHY-0705682; and by the University of Pittsburgh
Particle Physics, Astrophysics, and Cosmology Center. HJH thanks
CERN Theory Division for hospitality.



\begin{thebibliography}{99}
\baselineskip 18pt


\bibitem{LHC7-2011}
G.\ Aad {\it et al.,} [ATLAS Collaboration], Phys.\ Lett.\ B\,{\bf 710},
49 (2012) [arXiv:1202.1408 [hep-ex]];
S.\ Chatrchyan {\it et al.,} [CMS Collaboration],
Phys.\ Lett.\ B\,{\bf 710}, 26 (2012) [arXiv:1202.1488 [hep-ex]].


\bibitem{Csaki:2003dt}
  C.~Csaki, C.~Grojean, H.~Murayama, L.~Pilo and J.~Terning,
  Phys.\ Rev.\ D {\bf 69}, 055006 (2004)
  [hep-ph/0305237];
  C.~Csaki, C.~Grojean, L.~Pilo and J.~Terning,
  Phys.\ Rev.\ Lett.\  {\bf 92}, 101802 (2004)
  [hep-ph/0308038].


\bibitem{SekharChivukula:2001hz}
R.\ S.\ Chivukula, D.\ A.\ Dicus,  H.\ J.\ He, Phys.\ Lett.\ B
{\bf 525}, 175 (2002) [hep-ph/0111016];
R.\ S.\ Chivukula and H.\ J.\ He, Phys.\ Lett.\ B {\bf 532}, 121
(2002) [hep-ph/0201164];
R.\ S.\ Chivukula, D.\ A.\ Dicus, H.\ J.\ He, S.\ Nandi,
Phys.\ Lett.\ B {\bf 562}, 109 (2003) [hep-ph/0302263];
H.\ J.\ He, Int.\ J.\ Mod.\ Phys.\ A {\bf 20}, 3362 (2005) [hep-ph/0412113]
R.\ S.\ Chivukula, H.\ J.\ He, M.\ Kurachi, E.\ H.\ Simmons, M.\ Tanabashi, %
Phys.\ Rev.\ D {\bf 78}, 095003 (2008) [arXiv:0808.1682]


\bibitem{3site}
 R.\ S.\ Chivukula,  B.\ Coleppa, S.\ Di\,Chiara,
 E.\ H.\ Simmons, H.\ J.\ He, M.\ Kurachi,
 M.\ Tanabashi, Phys.\ Rev.\ D {\bf 74}, 075011 (2006) [arXiv:hep-ph/0607124].


\bibitem{Abe:2012fb}
 T.\ Abe, N.\ Chen, H.\ J.\ He, JHEP (2012), [arXiv:1207.4103 [hep-ph]].


\bibitem{July4}
G.\ Aad {\it et al.,} [ATLAS Collaboration],
Phys.\ Lett.\ B\,{\bf 716}, 1 (2012) [arXiv:1207.7214 [hep-ex]]; \\
S.\ Chatrchyan {\it et al.,} [CMS Collaboration],
Phys.\ Lett.\ B\,{\bf 716}, 30 (2012) [arXiv:1207.7235 [hep-ex]].


\bibitem{SekharChivukula:2005xm}
  R.~S.~Chivukula, E.~H.~Simmons, H.~J.~He, M.~Kurachi and M.~Tanabashi,
  Phys.\ Rev.\  D {\bf 72}, 015008 (2005)
  [arXiv:hep-ph/0504114].


\bibitem{Cacciapaglia:2004rb}
  G.~Cacciapaglia, C.~Csaki, C.~Grojean and J.~Terning,
  Phys.\ Rev.\  D {\bf 71}, 035015 (2005)
  [arXiv:hep-ph/0409126].


\bibitem{Foadi:2004ps}
  R.~Foadi, S.~Gopalakrishna and C.~Schmidt,
  Phys.\ Lett.\  B {\bf 606}, 157 (2005)
  [arXiv:hep-ph/0409266].


\bibitem{Chivukula:2005bn}
R.~S.~Chivukula, E.~H.~Simmons, H.~J.~He, M.~Kurachi and M.~Tanabashi,
Phys.\ Rev.\ D {\bf 71}, 115001 (2005)
[arXiv:hep-ph/0502162].


\bibitem{Casalbuoni:2005rs}
R.~Casalbuoni, S.~De Curtis, D.~Dolce and D.~Dominici,
Phys.\ Rev.\ D {\bf 71}, 075015 (2005)
[arXiv:hep-ph/0502209].


\bibitem{tri-3site}
 A.\ Belyaev, R.\ S.\ Chivukula, N.\ D.\ Christensen,
 H.\ J.\ He, M.\ Kurachi, E.\ H.\ Simmons, M.\ Tanabashi,
 Phys.\ Rev.\ D {\bf 80}, 055022 (2009) [arXiv:0907.2662];
 and presentation in the Proceedings of International Workshop on
 ``Strong Coupling Gauge Theories in LHC Era" (SCGT-2009),
 arXiv:1003.1786 [hep-ph].


\bibitem{PRD2008}
 H.\ J.\ He, Y.\ P.\ Kuang, Y.\ Qi, B.\ Zhang,
 A.\ Belyaev, R.\ S.\ Chivukula, N.\ D.\ Christensen,
 A.\ Pukhov, E.\ H.\ Simmons,
 Phys.\ Rev.\ D {\bf 78}, 031701 (2008) [arXiv:0708.2588].


\bibitem{3siteLHC-ex}
 T.\ Ohl and C.\ Speckner,
 Phys.\ Rev.\ D {\bf 78} (2008) 095008 [arXiv:0809.0023];
 T.\ Abe, T.\ Masubuchi, S.\ Asai, and J.\ Tanaka,
 Phys.\ Rev.\ D {\bf 84} (2011) 055005 [arXiv:1103.3579];
 F.\ Bach and T.\ Ohl,
 Phys.\ Rev.\ D {\bf 85} (2012) 015002 [arXiv:1111.1551].


\bibitem{Casalbuoni:1985kq}
  R.~Casalbuoni, S.~De Curtis, D.~Dominici and R.~Gatto,
  Phys.\ Lett.\  B {\bf 155}, 95 (1985).


\bibitem{Casalbuoni:1996qt}
R.~Casalbuoni {\em et al.},
{Phys. Rev.} {\bf D53}, 5201 (1996) [hep-ph/9510431].


\bibitem{Bando:1985ej}
  M.~Bando, T.~Kugo, S.~Uehara, K.~Yamawaki and T.~Yanagida,
  Phys.\ Rev.\ Lett.\  {\bf 54}, 1215 (1985).


\bibitem{Bando:1985rf}
  M.~Bando, T.~Kugo and K.~Yamawaki,
  Nucl.\ Phys.\  B {\bf 259}, 493 (1985).


\bibitem{Bando:1988ym}
M.~Bando, T.~Fujiwara, and K.~Yamawaki,
Prog.\ Theor.\ Phys.\  {\bf 79}, 1140 (1988).

\bibitem{Bando:1988br}
M.~Bando, T.~Kugo, and K.~Yamawaki,
{Phys. Rept.} {\bf 164}, 217 (1988).

\bibitem{Harada:2003jx}
M.~Harada and K.~Yamawaki,
Phys.\ Rept.\ {\bf 381}, 1 (2003) [hep-ph/0302103].


\bibitem{SMuni}
J.\ M.\ Cornwall, D.\ N.\ Levin, and G.\ Tiktopoulos,
Phys.\ Rev.\ Lett.\ {\bf 30} (1973) 1268;
Phys.\ Rev.\ D\,{\bf 10} (1974) 1145;
C.\ H.\ Llewellyn Smith, Phys.\ Lett.\ {\bf 46}B (1973) 233.
D.\ A.\ Dicus and V.\ S.\ Mathur, Phys.\ Rev.\ D\,{\bf 7} (1973) 3111;
B.\ W.\ Lee, C.\ Quigg, and H.\ B.\ Thacker,
Phys.\ Rev.\ Lett.\ {\bf 38} (1977) 883;
Phys.\ Rev.\ D\,{\bf 16} (1977) 1519;
M.\ S.\ Chanowitz and M.\ K.\ Gaillard,
Nucl.\ Phys.\ B {\bf 261} (1985) 379.


\bibitem{SekharChivukula:2008mj}
R.\ S.\ Chivukula, H.\ J.\ He, M.\ Kurachi, E.\ H.\ Simmons, %
M.\ Tanabashi, Phys.\ Rev.\ D {\bf 78}, 095003 (2008)
[arXiv:0808.1682].


\bibitem{Christensen:2008py}
  N.~D.~Christensen and C.~Duhr,
  Comput.\ Phys.\ Commun.\  {\bf 180}, 1614 (2009)
  [arXiv:0806.4194 [hep-ph]].


\bibitem{Christensen:2009jx}
  N.~D.~Christensen, P.~de Aquino, C.~Degrande, C.~Duhr,
  B.~Fuks, M.~Herquet, F.~Maltoni and S.~Schumann,
  Eur.\ Phys.\ J.\ C {\bf 71}, 1541 (2011)
  [arXiv:0906.2474 [hep-ph]].


\bibitem{helicity}
K.\ Hagiwara and D.\ Zeppenfeld, DESY-85-133; %
K.\ Hagiwara, R.\ D.\ Peccei, D.\ Zeppenfeld and K.\ Hikasa,
DESY-86-058 (1986); K.\ Hagiwara and D.\ Zeppenfeld, %
KEK Preprint 87-158 (1988).


\bibitem{cteq6}
J. Pumplin, D.\ R.\ Stump, J. Huston, H.\ L.\ Lai, P. Nadolsky and
W.\ K.\ Tung, JHEP {\bf 07}, 012 (2002).


\bibitem{j-separation}
This kind of cut should be applied to separated jets.
Experimentally, two jets are separable if $\,\Delta R_{jj}>0.5$\,
[see for instance, S. Ask (ATLAS Collaboration), arXiv:1106.2061],
and two jet-cones do not overlap at all if $\,\Delta R_{jj}>1$\,.\,
Hence the cut (\ref{eq:DR-cut1}) can be realized.


\bibitem{Pukhov:1999gg}
  A.~Pukhov, E.~Boos, M.~Dubinin, V.~Edneral, V.~Ilyin, D.~Kovalenko,
  A.~Kryukov and V.~Savrin, {\it et al.},
  arXiv:hep-ph/9908288.


\bibitem{Pukhov:2004ca}
  A.~Pukhov,
  arXiv:hep-ph/0412191.

\bibitem{PDG}
K. Nakamura {\it et al.,} [Particle Data Group], %
J.\ Phys.\ G\,{\bf 37},  075021 (2010) [No.\,7A].


\end{thebibliography}
\end{document}

Title: Discovering New Gauge Bosons of Electroweak Symmetry Breaking at LHC-8
\\
Authors:  Chun Du, Hong-Jian He, Yu-Ping Kuang, Bin Zhang (Tsinghua),
Neil D. Christensen (Pittsburgh), R. Sekhar Chivukula, Elizabeth H. Simmons (MSU)
\\
Comments: PRD final version (only minor refinements showing the consistency with new LHC data),
11 pages, 5 Figs, 2 Tables
\\
We study the physics potential of the 8TeV LHC (LHC-8) to discover, during
its 2012 run, a large class of extended gauge models or extra dimensional
models whose low energy behavior is well represented by an SU(2)^2 x U(1)
gauge structure. We analyze this class of models and find that with a combined
integrated luminosity of 40-60/fb at the LHC-8, the first new Kaluza-Klein mode
of the W gauge boson can be discovered up to a mass of about 370-400 GeV,
when produced in association with a Z boson.
\\
Report number: CERN-PH-TH/2012-177, MSUHEP-120618, PITT-PACC-1206, TUHEP-TH-12177